\documentclass[preprint2]{aastex}

\usepackage{natbib}
\usepackage[english]{babel}

\slugcomment{To appear in The Astronomical Journal (AJ)}

\shorttitle{Ruling out the orbital decay of the WASP-43b}
\shortauthors{Hoyer et al.}

\begin{document}


\title{Ruling out the orbital decay of the WASP-43b exoplanet}


\author{Sergio Hoyer\altaffilmark{1,2}}
\email{shoyer@iac.es}

\author{Enric Pall{\'e} \altaffilmark{1,2}}
\author{Diana Dragomir \altaffilmark{3}}
\and
\author{Felipe Murgas \altaffilmark{4,5}}

\altaffiltext{1}{Instituto de Astrof{\'i}sica de Canarias, E-38205 La Laguna, Tenerife, Spain}
\altaffiltext{2}{Universidad de La Laguna, Dpto. Astrof{\'i}sica, E-38206 La Laguna, Tenerife, Spain}
\altaffiltext{3}{The Department of Astronomy and Astrophysics, University of Chicago, 5640 S Ellis Ave, Chicago, IL 60637, USA}
\altaffiltext{4}{Univer Grenoble Alpes, IPAG, F-38000 Grenoble, France}
\altaffiltext{5}{CNRS, IPAG, F-38000 Grenoble, France}

\begin{abstract}

We present 15 new transit observations of the exoplanet WASP-43b in
the $i'$,$g'$, and $R$ filters with the 1.0-m telescopes of Las
Cumbres Observatory Global Telescope (LCOGT) Network and the IAC80
telescope. We combine our 15 new light curves with 52 others from
literature, to analyze homogeneously all the available transit light
curves of this exoplanet.  By extending the time span of the
monitoring of the transits to more than $5~yr$, and by analyzing the
individual mid-times of 72 transits, we study the proposed shortening
of the orbital period of WASP-43b.  We estimate that the times of
transit are well-matched by our updated ephemeris equation, using a
constant orbital period.  We estimate an orbital period change rate no
larger than $\dot{P}=-0.02  \pm 6.6~ms~yr^{-1}$, which is fully
consistent with a constant period.  Based on the timing analysis, we
discard stellar tidal dissipation factors $Q_{*}<10^{5}$.  In
addition, with the modelling of the transits we update the system
parameters: $a/Rs=4.867(23)$, $i=82.11(10)^{\circ}$ and
$R_p/R_s=0.15942(41)$, noticing a difference in the relative size of
the planet between optical and NIR bands.

\end{abstract}



\bibliographystyle{apj}

\section{Introduction}

Exoplanets with ultra short orbital periods ($\lesssim 1~ d$) are
rare.  Of the current $\sim2000$ confirmed exoplanets, 17 objects
exhibit orbital periods of less than 1.3 days \citep[based on the
  exoplanets.org public compilation:][]{Han_2014}.  The most extreme
case of these objects is the Earth-size planet Kepler-178b, with an
orbital period of only 8.5 ~h \citep{Sanchis_Ojeda_2013}.  Also, 8/17
of these planets have radii of $\leq 2 ~R_{Earth}$, while the rest
spans between 10 and 20 Earth radii with masses no larger than than
$2~M_{Jupiter}$, with the exception of WASP-18b which has an estimated
mass of $10~M_{Jupiter}$ \citep{Hellier_2009}.  One of these objects
is WASP-43b \citep{Hellier_2011}, a $0.93~R_{Jupiter}$ size planet
with a mass of $1.83~M_{Jupiter}$, orbiting a K star in only $0.81~d$
\citep{Gillon_2012}.  Since its detection and confirmation by the
Wide-Angle Search for Planets (WASP) group, this exoplanet has been
studied intensively.  For example, \citet{Gillon_2012} reported an
extended follow-up which included the observations of 20 transits with
the $0.6~m$ TRAnsiting Planets and PlanetesImals Small Telescope
(TRAPPIST) in a $I+z$ filter, 3 transits with the Euler telescope
($Gunn-r'$ band) and 7 secondary transits in the NIR (5 in the
$Sloan-z'$ filter with TRAPPIST and 2 in narrow-band filters at the
ESO Very Large Telescope). This work presented an estimation of the
planet temperature, $T_{eq}=1440~K$, and a re-estimation the planet
mass, $2~M_{Jupiter}$. Shortly after, \citet{Wang_2013} reported two
secondary transit observations, deriving a $T_{planet}\sim 1850~K$.
Later, \citet{Blecic_2014} also observed secondary transits with the
Spitzer space telescope. Based on the central times of
\citet{Gillon_2012} transits and amateur data from the {\it Exoplanet
  Transit Database} \citep{Poddan__2010}, \citet{Blecic_2014} reported
for the first time a hint of change in the orbital period of
$\dot{P}=-0.095\pm0.036 ~s~yr^{-1}$.  Due to the closeness to its host
star and the relative high stellar brightness ($V=12.4 ~mag$),
WASP-43b is a very suitable target for atmospheric studies.  One
example of these studies is the transmission spectrum observations
obtained with OSIRIS instrument at the 10.4-m Gran Telescopio Canarias
(GTC) presented by \citet{Murgas_2014}. In addition to the tentative
detection of $NaI$ in the atmosphere of WASP-43b, \citet{Murgas_2014}
re-estimated the shortening of the orbital period to $\dot{P}=-0.15\pm 0.06~s~yr^{-1}$ 
by including 5 new epochs in the analysis.
\citet{Chen_2014} presented 7 broad-band simultaneous observations
(from $g'$ to K filter) of a primary and a secondary transit using
GROND instrument at the 2.2m MPG/ESO Telescope, reporting a possible
difference between the transit depths of the optical and NIR light
curves.  Furthermore, using a re-analysis of \citet{Gillon_2012} and
amateur data, \citet{Chen_2014} also calculated a non-negligible
$\dot{P}=-0.09\pm 0.04~s~yr^{-1}$.  Recently, 6 transits and 5
eclipses observations with WFC3 instrument on board of the Hubble
Space Telescope (HST) were presented to constrain the water abundance
in the WASP-43b atmosphere \citep{Kreidberg_2014}.  These data were
revisited by \citet{Stevenson_2014} and \cite{Kataria_2015} to map the
thermal structure of the planet along the full orbital phase and to
fit a circulation model of the planet, respectively.  Along with these
studies on the exoplanet atmosphere, \citet{Stevenson_2014} reported
high precision mid-times for each transit.  Additional transits were
also reported by \citet{Macie_2013}, 1 epoch in $R$ and 1 without
filter, and \citet{Ricci_2015}, 7 observations of 6 different epochs
in the $R$, $V$ and $i'$ filters, giving a a new value of
$\dot{P}=-0.03 \pm 0.03 ~s~yr^{-1}$. During the writing of this work,
9 new transits observations of WASP-43b were reported by
\citet{Jiang_2015}.  Combining their data to the literature transits,
they found that the amplitude of orbital decay is consistent with
previous works, $\dot{P}=-0.029 \pm 0.008~s~yr^{-1}$ and therefore, a
slow decreasing rate of the period was not discarded.

Here, we present a total of 15 new transit observations on 9 different
epochs.  Of these, 14 transits were obtained with Las Cumbres
Observatory Global Telescope (LCOGT) Network while one transit
was observed with the 0.8m IAC80 Telescope.  Combining these new data
and all the literature transits available for WASP-43b, we perform an
homogeneous analysis of the light curves, with special focus in the
transit times, to probe the proposed orbital decay. Along with the
timing studies, we also revisited the system and transit parameters of
WASP-43b. In Section \ref{data} we present our observations, in
Section \ref{modelling} we describe the modelling of the transits, in
Section \ref{timing} we present the timing analysis of the transits
divided in two parts: in Section \ref{timing-a} we present the time
analysis of the 58 transits analyzed in this work while in Section
\ref{timing-b} we include into the analysis additional timing
information available for this planet. In Section \ref{Q} we use our
timing results to constrain the tidal dissipation efficiency of
WASP-43. Finally in Section \ref{conclusions} we present our
conclusions.

\section{Observations}
\label{data}

We present 15 new observations, obtained in 9 different transit
epochs, of the exoplanet WASP-43b.  Of these, 6 events were observed
simultaneously in the $i'$ and $g'$ filters with the 1.0-m LCOGT
Network, other two transits were observed only with the $g'$ or the
$i'$ filter also at the 1.0-m LCOGT, and one additional epoch was
observed in the $R$ band with the 0.8-m IAC80 telescope at Teide
Observatory, Spain.  The node of the LCOGT Network located at Cerro
Tololo Interamerican Observatory (CTIO) in Chile was used to observed
6 of the epochs.  The 1.0-m telescopes at the CTIO site are equipped
with Sinistro cameras, with a Field-of-View (FoV) of $27\times27
~arcmin^{2}$, a pixel size of $0.389~arcsec$, and a readout time of
$51~s$ without binning.  Other two epochs were observed using the
telescopes of the node at the South African Astronomical Observatory
(SAAO).  This node was equipped with a Sbig camera with a FoV of
$16\times16 ~arcmin^{2}$, a pixel scale of 0.232 $arcsec$, and a
readout time of $15.5~s$ when using the $2\times2$ binning mode. Some
of the observations were taken with the telescopes defocused in order
to increase the number of counts per object without reaching the
non-linearity regime of the detectors. The egress of the transit in
2014-May-26 was lost due to bad weather conditions at CTIO.

The transit at IAC80 telescope was observed using the CAMELOT
instrument.  CAMELOT has a collector area of 2048x2048 $pixels^{2}$
with two readout amplifier channels, a FoV of 10.4x10.4 $arcmin^{2}$
and a pixel scale of 0.304 $arcsec$.  The observation was performed
using the 500 KHz readout speed in both channels without binning,
having thus a readout time of 4.4 $~s$.  Our last observation was
executed in 2016-Feb-04 and corresponds to the transit epoch 2329
(using as E=0 the reference epoch from \citet{Hellier_2011}).  With
this observation we extend the monitoring of the WASP-43b transits to
$5.2 ~yr$. The observing log of each observation is shown in Table
\ref{tab:table_obs}.

\subsection{Reduction and Photometry}

All images were processed using the pipeline described in
\citet{Bro13}. Briefly, the entirely automated procedure includes
bad-pixel masking, bias subtraction, dark subtraction, flat field
correction and astrometric solution. We used custom made Python
pipelines to perform differential aperture photometry on the target
and several stars in the FoV are used as reference. We choose the best
reference stars by identifying the objects which produce the lowest
RMS in the out-of-transit (oot) data.  The size of the aperture and
the ring used to measure the sky background were also chosen with the
same RMS criteria.  The final aperture radii were on the range of the
13-20 pixels, and the radii and widths of the sky rings were between
18-25 and 10-20 pixels, respectively.

\subsection{Literature transits}

We have included in our homogeneous analysis all the light curves
available in the literature.  We used 23 transits from
\citet{Gillon_2012}, 2 transits from \citet{Macie_2013}, the 7 light
curves from \citet{Chen_2014} of the epoch=499 transit, the 5 transits
from \citet{Murgas_2014} and the 7 light curves from
\citet{Ricci_2015} (obtained in 6 different epochs).  The epoch and
filter of each transit is shown in Table \ref{tab:results}.  We also
fit the 8 transits recently reported \citet{Jiang_2015} in order to
include their epochs in our timing analysis although we did not use it
in Section \ref{modelling} to derive the final system parameters.

\begin{deluxetable}{cccccccc}[ht]
\tablewidth{0pt} 
\tabletypesize{\footnotesize}
\tablecaption{  Log of the observations.  \label{tab:table_obs} }

\tablehead{ 
\colhead{ Date \tablenotemark{a} }  & 
\colhead{Epoch \tablenotemark{b} }  & 
\colhead{Site}  & 
\colhead{Camera / }  & 
\colhead{Filter}&
\colhead{Exposure} & 
\colhead{Binning } & 
\colhead{Airmass}   \\
\colhead{ (yyyymmdd)  }  &
\colhead{}      &
\colhead{}    &
\colhead{Intrum. ID} &
\colhead{}     &  
\colhead{(s)}  &
\colhead{(pix $\times$ pix) } }
 
\startdata 
20140501 & 1536  & SAAO  & Sbig / kb75     & $g'$      & 30            & 2x2     & 1.1-1.82      \\
   &       & SAAO  & Sbig / kb70     & $i'$      & 30            & 2x2     & 1.1-1.64      \\
20140509 & 1546  & CTIO  & Sinistro / fl04 & g      & 60            & 1x1     & 1.1-1.06-1.14 \\
      &       & CTIO  & Sinistro / fl03 & $i'$      & 30            & 1x1     & ''            \\
20140513 & 1551  & CTIO  & Sinistro / fl03 & g      & 45            & 1x1     & 1.07-1.37     \\
      &        & CTIO  & Sinistro / fl02 & $i'$      & 40            & 1x1     &        ''       \\
20140526 & 1567  & CTIO  & Sinistro / fl04 & $i'$      & 60            & 1x1     & 1.09-1.61     \\
       &      & CTIO  & Sinistro / fl03 & $g'$      & 70            & 1x1     & ''           \\
20140626 & 1605  & CTIO  & Sinistro / fl04 & $g'$      & 45            & 1x1     & 1.28-1.89     \\
20141229 & 1834  & CTIO  & Sinistro / fl04 & $g'$      & 60            & 1x1     & 1.67-1.10     \\
      &       & CTIO  & Sinistro / fl03 & $i'$      & 55            & 1x1     & ''           \\
20150211 & 1888  & CTIO  & Sinistro / fl04 & $g'$      & 65            & 1x1     & 1.24-1.06     \\
      &       & CTIO  & Sinistro / fl03 & $i'$      & 55            & 1x1     &''             \\
20150502 & 1986  & Teide & CAMELOT         & $R$      & 50            & 1x1     & 1.27-1.99    \\
20160204 & 2329  & SAAO  & Sbig / kb76     &  $i'$      & 30            & 2x2     & 1.29-1.08      \\
\enddata

\tablenotetext{a}{ corresponds to the date stamp in the header of the
  first frame of the run.}  \tablenotetext{b}{ transit epoch of the
  observation, here E=0 corresponds to the epoch reported by
  \cite{Hellier_2011} }

\end{deluxetable}

\begin{figure*}
\begin{center}
\includegraphics[width=2\columnwidth]{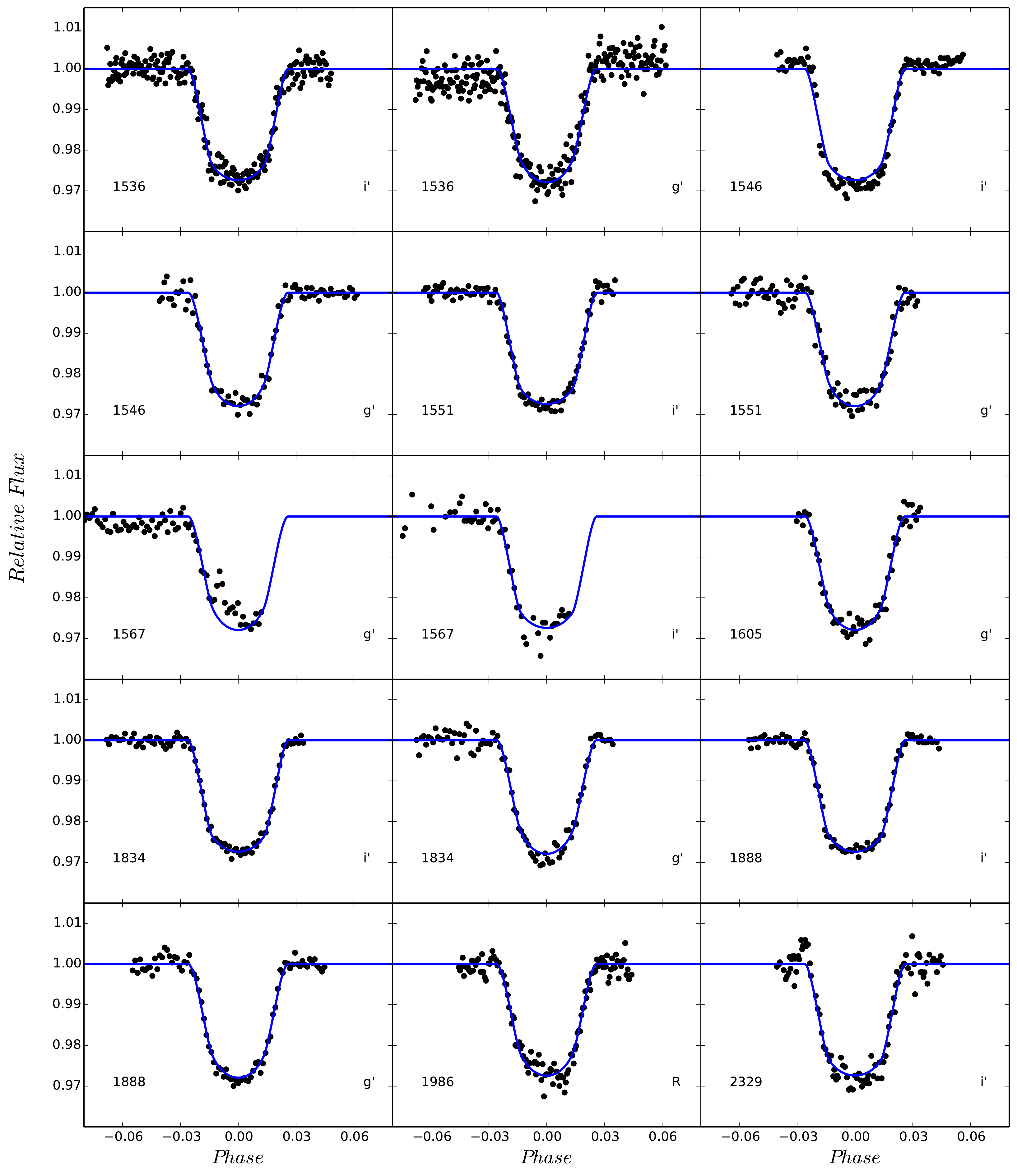}
\caption{The 15 new light curves presented in this work (black points)
  and the best-fitting model obtained in Section \ref{rprs_free} (blue
  solid line). The epoch and the filter of the transit observation is
  shown on the left and right bottom corner of each panel,
  respectively.  All the light curves were obtained with LCOGT Network
  except for the transit E=1986 which was observed with the IAC80
  Telescope.  \label{lcogt_lcs} }
\end{center}
\end{figure*}

\begin{figure*}
\begin{center}
\includegraphics[width=2\columnwidth]{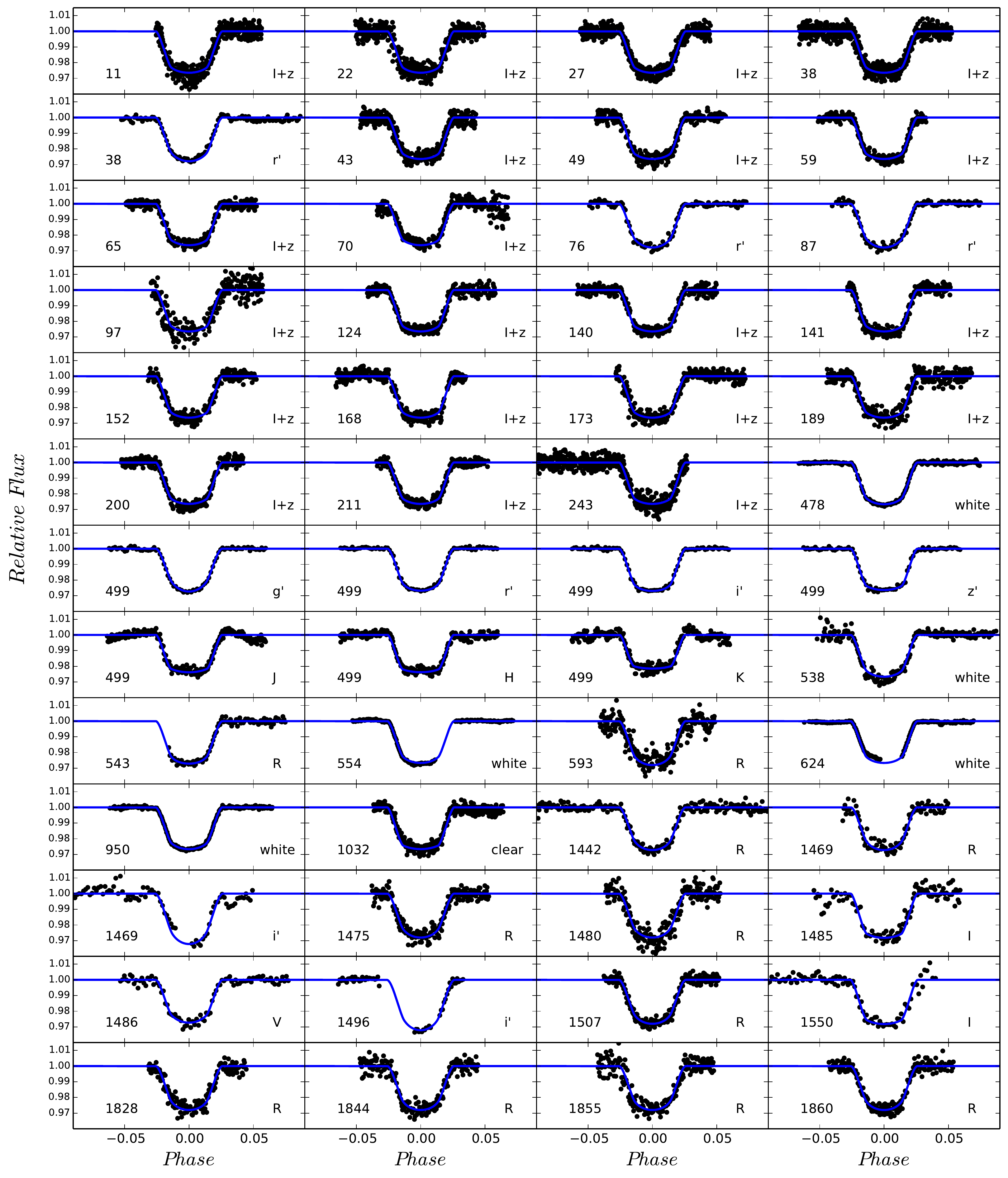}
\caption{We show all the literature transit light curves analyzed in
  this work and the best model we obtained in Section \ref{rprs_free}.
  The epoch and filter of the transit observation is shown on the left
  and right bottom corner of each panel,
  respectively. \label{literature_lcs}}
\end{center}
\end{figure*}

\section{Modelling}
\label{modelling}
We used the \textit{TAP} package \citep{Gazak_2012} for the
simultaneous modelling of all the light curves. This package allows to
fit for orbital and transit parameters using the analytic function of
\citet{Mandel_2002} to describe the exoplanetary transits.  TAP also
incorporates the wavelet method of \citet{Carter_2009} to estimate the
correlated noise in the light curves and the Markov Chain Monte Carlo
(MCMC) approach to calculate the uncertainties of the fitted
parameters.  In particular the parameters subject to be fitted are:
orbital period ($P$), inclination ($i$), eccentrycity ($e$), longitude
of the periastron ($w$), relative distance to the host star ($a/R_s$),
planet-to-star radii ratio ($R_p/R_s$), transit mid-time ($T_c$) and
the coefficients of a limb-darkening quadratic law ($u_1$ and $u_2$).
Furthermore, this version also allows to fit for a time dependent
linear function (i.e. $F_{slope}$ and $F_{shift}$) in addition to the
noise parameters ($\sigma_{red}$ and $\sigma_{white}$) from the
wavelet method, assuming that the correlated noise can be described by
a $1/f^{\gamma}$ function, where $f$ is the frequency and $\gamma$ is
assumed to be equal to 1 (see \citet{Carter_2009} for details). Two
different methodologies were employed.

\begin{figure}[h!]
\begin{center}
\includegraphics[width=.95\columnwidth]{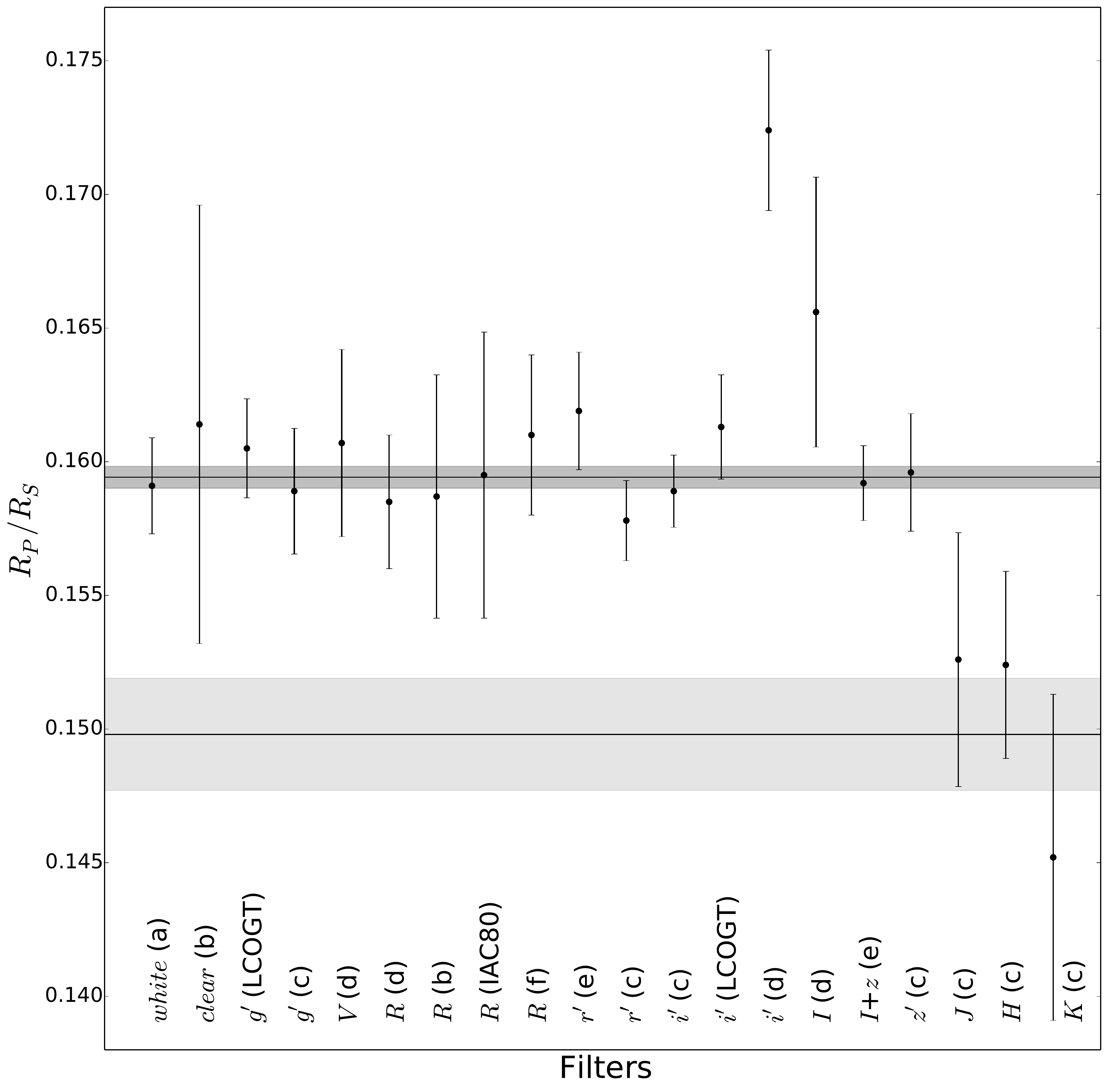} \figcaption{Here
  we show the values of $R_p/R_s$ obtained for each set of light
  curves modelled independently.  We label each {\it set} using the
  filter and author as follow: (a) \citet{Murgas_2014}, (b)
  \citet{Macie_2013}, (c) \citet{Chen_2014}, (d) \citet{Ricci_2015},
  (e) \citet{Gillon_2012}, (f) \cite{Jiang_2015} and (LCOGT) and
  (IAC80) are from this work.  The top black solid line and the grey
  region correspond to the resulting value, and its errors, of the
  joint modelling of all the light curves (Section \ref{RpRs_simul}).
  We unlinked from the joint modelling the $i'$ and $I$ from (d),
  which deviate noticeably from the rest of the sets (see text for
  details). The lower flat line shows the result from the joint
  modelling of the NIR transits: $J$, $H$ and $K$ from
  (c).  \label{fig:Kvsfilter} }
\end{center}
\end{figure}

\subsection{Method 1: Joint modelling with free $R_p/R_s$}
\label{rprs_free}
In order to explore any variation as a function of wavelength, in a
first step we grouped the transits according to the filter and
telescope and fit for $R_p/R_s$ and ($u_{1}$,$u_{2}$) on each set of
light curves. We modelled $i$ and $a/R_s$ simultaneously in all the
transits, fixing $P$ to $0.81347385~d$ from \cite{Murgas_2014}, and
$e=w=0$. The midtime of each transit (\,$T_c$) and the rest of the
parameters were left free to vary individually on each transit.  We
used 10 MCMC chains of $10^{5}$ links each.  The resulting parameters
were obtained from the median of the posterior distributions and its
respective uncertainties from the 16 and 84 percentile levels.  In
Fig. \ref{lcogt_lcs} and \ref{literature_lcs}, we show each transit
light curve and its final model for the LCOGT and literature transits,
respectively, and in Table \ref{tab:results} we show the parameters
obtained using \textit{TAP}. In Figure \ref{fig:Kvsfilter} we show the
values of $R_p/R_s$ for each group of light curves, which present a
very good agreement for the optical filters.  Only the two transits in
$I$ and $i'$ of \citet{Ricci_2015} deviate from the rest of the optical
band transits, probably due to the low quality of these light curves.
We also notice that the NIR transits depths ($J$-,$H$- and $K$- bands)
also deviate significantly from the optical values.  \citet{Chen_2014}
reported a similar deviation in the $z'$-,$H$- and $K$- bands, but
their differences were not statistically significant.  A multi-epoch
study in NIR bands would be needed to confirm this finding.

\subsection{Method 2: Joint modelling with common $R_p/R_s$}
\label{RpRs_simul}
To take advantage of the large amount of data we have, we use all the
light curves in the modelling to refine the final planet parameters.
We repeated the method described in Section \ref{rprs_free} but this
time fitting $R_p/R_s$ simultaneously for all the optical and the NIR
transits from \citet{Chen_2014}. Here we remove some transits from the
joint modelling since we have identified in the previous step that its
respective $R_p/R_s$ values deviates considerably from the rest of the
transits. In particular, we removed from the simultaneous modelling of
$R_p/R_s$ the $i'$ and $I$ transits from \citet{Ricci_2015}, and the
E=538, 554 and 624 transits from \citet{Murgas_2014} whose transit
depths were strongly affected by bad weather and/or technical issues
during observation.  We have not included in this step the data from
\citet{Jiang_2015} since these transits were only available at a final
stage of this work. As in method 1, the $u_1$ and $u_2$ coefficients
are fitted by groups.

By doing this we obtained $Rp/Rs=0.15942 \pm 0.00041$ and
$R_p/R_s=0.1498 \pm 0.0021$ for the optical and NIR bands,
respectively.  Additionally, we obtained $i=82.11\pm0.10$, and
$a/R_s=4.867^{+0.023}_{-0.025}$ from all the light curves.  As
expected, the values of $a/R_s$ and $i$ do not differ from those
obtained before, but the \textit{optical} $R_p/R_s$ is one order of
magnitude more precise by using the joint modelling.  Regarding the
mid-times of each transit we obtained values fully consistent between
the two methods.  Our derived $R_p/R_s$ is also consistent with the
values reported by \citet{Murgas_2014}, \citet{Stevenson_2014} and
\citet{Jiang_2015}: $0.15988^{+0.00133}_{-0.00145}$, $0.15948 \pm
0.00004$, and $0.15929 \pm 0.00045$, respectively.  On the other hand,
the $R_p/R_s$ derived from NIR transits deviates significatively from
these measurements (see Figure \ref{fig:Kvsfilter}), even though
\citet{Stevenson_2014} value is based on data which cover the $J$- and
$H$- filters.

\section{Timing Analysis} \label{timing}

For the modelling of our light curves we use the mid-exposure time of
each frame recorded in Julian Days (UTC). Later on, we transform the
resulting $T_c$ for each transit to Barycentric Julian Days (BJD) in
the Barycentric Dynamical Time (TDB) standard as suggested by
\citet{Eastman_2010}.  For the literature values we also performed the
necessary transformation of the central times of their original time
standard to $BJD_{TDB}$.  As mentioned before, we have obtained
simultaneous observations of 6 transits epochs with two 1-m LCOGT
telescopes (3 additional transits were not observed simultaneously).
The mid-times we obtained for each of the simultaneous observations
are consistent within the errors, with the exception of the transits
of epoch 1834, which deviates by $1.5\sigma$ (very likely due to an
underestimated error and/or unaccounted systematics in the $g'$
transit).  Furthermore, our retrieved mid-times for the literature
data are also consistent with the original reported values.  The
average difference between those mid-times is only $\sim0.35\sigma$.

\subsection{Timing results from the mid-times of our homogeneous analysis}
\label{timing-a}

To adjust the mid-times of the 67 transits obtained in Section
\ref{modelling}, we used a linear ephemeris equation of the form:
\begin{equation}
T(E) = T_0 + E \times P,
\label{eqn:general_linear}
\end{equation}
where the central time of the transit $T_{c}$ at epoch $E$, is
calculated with respect to the reference time $T_0$ using the orbital
period $P$.  To update the linear ephemeris we used the
\verb|scipy.optimize| module of Python and the \verb|emcee| MCMC
sampler implementation \citep{Foreman_Mackey_2013}.  We fit for $P$
and $T_0$, and estimate its uncertainties using 1000 walkers with 5000
links each.  In a first step, we identified that the two transits of
epoch 1469 deviate by more than $147~s$. As these transits were not
included in the analysis on the original publication (private
communication with the author) we also decided to remove them from the
timing analysis.   After repeating the MCMC analysis for the rest of
the 65 transits, we obtained the following ephemeris equation:
\small
\begin{equation}
T(E) = 2455528.868602(47) + E \times 0.813474077(54),
\label{eqn:my_linear}
\end{equation}
\normalsize
where the quoted values and their uncertainties (in parenthesis) were
adopted from the 50\%, 16\% and 84\% percentiles of the drawn
posterior distributions, respectively.  The RMS of the timing residuals is $37~s$.
The $\chi^2$ of this fit is 121, the reduced-$\chi^2$ is 1.93 and the
Bayesian Information Criterion ($BIC=2log(\nu)+\chi^2$, where $\nu$ is
the number of degrees of freedom) is 130.

To probe for orbital decay, i.e. a shortening of the orbital period
with time, we also explore the possibility that the transit times can
be adjusted by including a quadratic additional term. Therefore we fit
the transit times using, as in \citet{Adams_2010}, the following
ephemeris equation:
\begin{equation}
T_{c}(E) = T_{0} + E \times P + \frac{1}{2} \delta P \times E(E-1),
\label{eqn:general_non_linear}
\end{equation}
where the new term $\delta P$ corresponds to the change rate of the
orbital period per $epoch^{-2}$ ($\delta P= \dot{P} \times P$).  Using
the same procedure as in the linear case, the resulting values of this
quadratic fit are: $\delta P = (0.38 \pm 1.90) \times 10^{-10} ~d
~epoch^{-2}$, $P=0.81347404(19)~d$ and $T_0(BJD_{TDB}) =
2455528.868610(66)$.  This new $\delta P$, which can be translated to
$\dot{P}=1.5 \pm 7.3~ms ~yr^{-1}$, is fully consistent with a constant
orbital period.

This quadratic fit has a $RMS$ of $37~s$, a $\chi^2$ of 121, a
reduced-$\chi^2$ of 1.96 and a $BIC_{quad}=134$.  These numbers
suggest that the linear equation is the most likely function to
describe the transits of WASP-43b.  

\subsection{Including additional transit mid-times}
\label{timing-b}

\begin{figure*}[ht]
\begin{center}
\includegraphics[width=1.75\columnwidth]{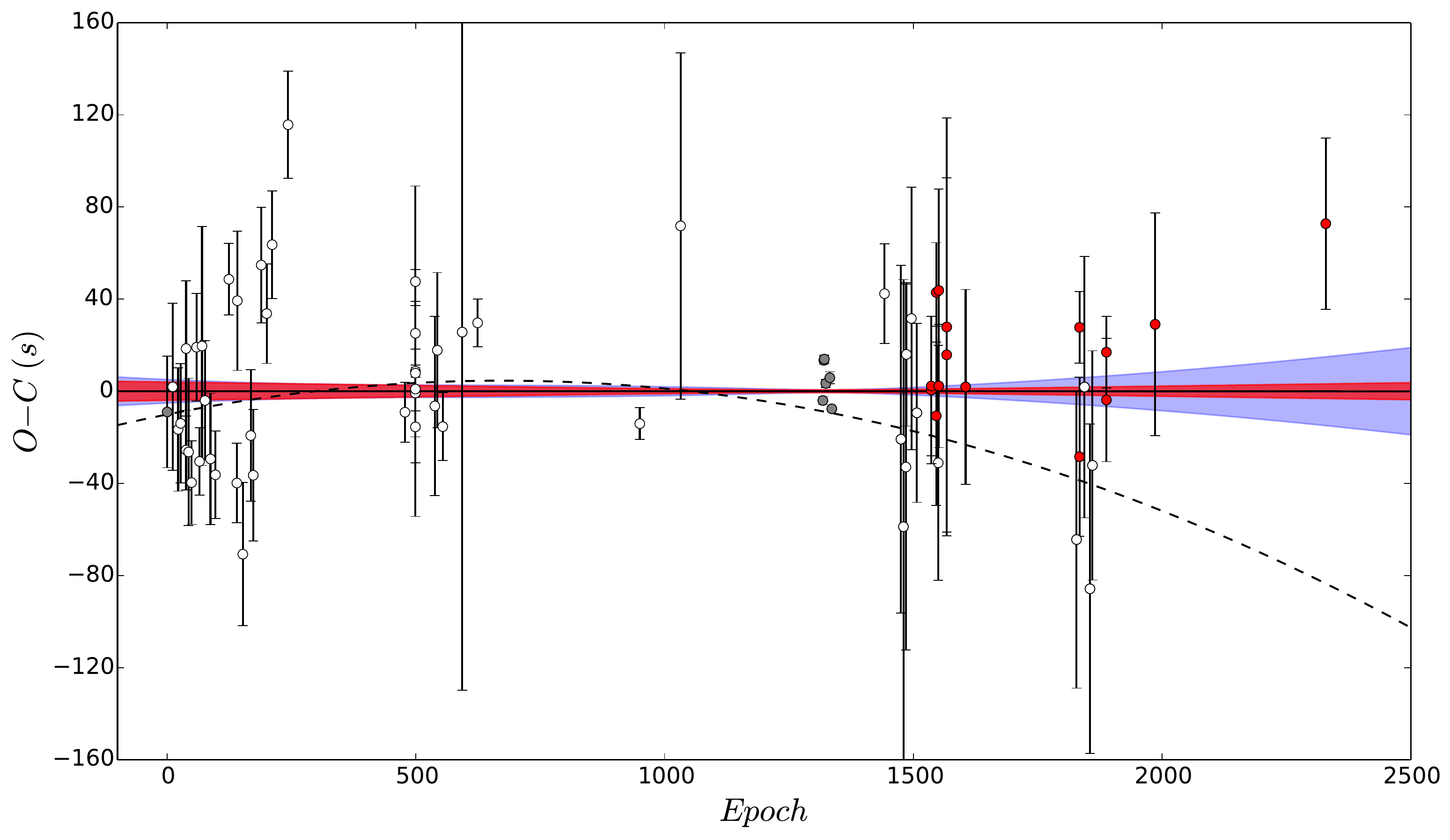}
\caption{\textit{Observed minus Calculated} diagram of 72 transit
  mid-times of WASP-43b.  The timing residuals obtained from the 65 light
  curves analyzed in this work are shown. The white and
  red points correspond to the literature and the new transits
  presented in this work, respectively, while the gray points
  represent the mid-times reported by \citet{Hellier_2011} and
  \citet{Stevenson_2014}. The timing residuals are based on our
  updated linear ephemeris equation (Eq. \ref{eqn:all_linear}), and
  the red region represents its $\pm1\sigma$ errors. The solid black
  line and the light blue region represent the fitted quadratic
  function and its uncertainties, respectively. The dashed line
  corresponds to the changing period function reported by
  \citet{Jiang_2015}.  \label{fig:o-c}}
\end{center}
\end{figure*}

Although we could not include in our modelling the transit light
curves presented in \citet{Hellier_2011} and \citet{Stevenson_2014},
we check our results by using their published mid-times.  Since we do
not have any common transit as a comparison point we use their
reported uncertainties without scaling them.  Thus, we repeat the
fitting process using a total of 72 transits, resulting in the most
comprehensive study of the WASP-43b transits to date.  We obtain in
this case a $\dot{P} =-0.02 \pm 6.6 ~ms ~yr^{-1}$, consistent again with
the constant orbital period case. Furthermore, in this case the BIC
also favors the linear equation ($BIC_{lin}=353$) over a non-constant
period scenario ($BIC_{quad}=358$).  Therefore, this new value of $\dot{P}$ rules out definitively a rapid
or slowly period decay of the WASP-43b exoplanet.  Is worth mentioning
that our monitoring includes a transit in the epoch=2329, which
extends the monitoring $381~d$ beyond the last epoch reported by
\citet{Jiang_2015}, totaling more than 5 years of follow-up.  Long
term monitoring is critical to probe small amplitude timing variations
in the transits of exoplanets \citep[e.g. see Fig.14  of][]{Birkby_2014,Hoyer_2016}.

Finally, using the 72 transits we calculate the updated linear
ephemeris function:
\small
\begin{equation}
T(E) = 2455528.868634(46) + E \times 0.813473978(35) ,
\label{eqn:all_linear}
\end{equation}
\normalsize
where the $RMS$ of the timing residuals is $35~s$, the reduced-$\chi^{2} =
4.9$ and the $BIC=353$.


The \textit{Observed minus Calculated} diagram of the transit
mid-times based on this linear fit is shown in Figure \ref{fig:o-c},
where the open points represent the literature transits analyzed in
Section \ref{modelling}, the gray points are from \cite{Hellier_2011}
and \citet{Stevenson_2014}, and the red points correspond to the new
transits presented in this work. The red region corresponds to the
$1\sigma$ uncertainties of the linear fit.  In this Figure the fitted
quadratic function based on the 72 transits and its $1\sigma$ errors
are represented by the solid black line and the blue light region,
respectively.  For comparison we plot with the dashed line the decay
function obtained by \citet{Jiang_2015}.  It is clear that despite the
errors in the mid-times and the dispersion of the residuals, the most
recent epochs do not follow the proposed  $\dot{P}=-28 ~ms ~yr ^{-1}$.  

\section{Constraints on Q*}
\label{Q}

Stellar tidal dissipation, usually characterized through the quality
or efficiency factor, $Q_{*}$, can drives the planet orbital motion
and determine, for example, the life time before it collides with its
host star \citep{Levrard_2009, Penev_2012}.  Therefore, empirical
constraints of $Q_{*}$ are relevant for models of formation and
evolution of close-in exoplanets.  Measuring the shortening of orbital
periods of transiting exoplanets is a direct way to estimate $Q_{*}$
\citep[e.g.][]{Matsu_2010,Penev_2012, Birkby_2014,Hoyer_2016}.  For
WASP-43, previous studies pointed to $Q_{*}$ between $10^{4}-10^{10}$
\citep{Hellier_2009, Blecic_2014, Jiang_2015}.  In particular, using
the equations in \cite{Birkby_2014}, we calculate that the previous
reported $\dot{P} =-30~ms~yr^{-1}$ implies a $Q_{*}\approx 4\times
10^{4}$ and therefore a $T_{shift} \approx 170~s$ in the time arrival
of the transits after 5 years. Such large time deviation is not
detected in our $O-C$ diagram (Figure \ref{fig:o-c}).  Now, based on
the uncertainties of our estimation of $\dot{P}$
($\pm6.6~ms~yr^{-1}$), we derive a $Q_{*}\approx 2\times10^{5}$ and a
$T_{shift}\approx 39 ~s$ after $5~yr$ which is is consistent with our
transit timing, considering the $RMS=35~s$ of the time residuals we
obtained.  Larger values of $Q_{*}$ result, after $5~yr$, in time
deviations of smaller amplitude than the dispersion of the
$O-C$. Thus, for WASP-43 we can discard $Q_{*}$ values smaller than
$10^{5}$.  With this precision in $\dot{P}$, and if the dispersion of
residuals of the transit times continues in the $30~s$ level after 5
additional years (i.e. around the epoch 4500), it would be
possible to probe the $Q_{*}<10^{6}$ limit. 

\section{Conclusions}
\label{conclusions}

With the timing analysis of 72 transit of WASP-43b, we have estimated
that the change rate of the orbital period is consistent with zero.
By extending the time span of the monitoring to more than $5~yr$,
i.e. to 2329 orbits, we have constrained the orbital decay to
$\dot{P}=-0.02\pm6.6~ms~yr^{-1}$, which is 3 order smaller than
values previously reported.  Based on our estimation of $\dot{P}$, we
can discard $Q_{*} < 10^{5}$ for WASP-43.  Finally, by fitting
together the optical transit light curves we have a re-estimation of
the relative size of the planet, $R_p/R_s=0.15942\pm0.00041$, a value
also consistent with previous studies.  We also found a difference between
the $R_p/R_s$ derived from optical and NIR filters.  We note that the
HST data of \citet{Stevenson_2014} cover the $1.1-1.7 \mu m$
wavelength range, i.e, it includes part of the $J$ and $H$ bands, and
their reported $R_p/R_s$ is fully consistent with our \textit{optical}
results. Thus, additional NIR observations are needed to state
definitively this discrepancy.

\acknowledgments 

This work makes use of observations made in LCOGT network and the
IAC80 telescope operated on the island of Tenerife by the IAC in the
Spanish Observatorio del Teide.  SH acknowledges financial support
from the Spanish Ministry of Economy and Competitiveness (MINECO)
under the 2011 Severo Ochoa Program MINECO SEV-2011-0187.  This work
is partly financed by the Spanish Ministry of Economics and
Competitiveness through projects ESP2013-48391-C4-2-R and
ESP2014-57495-C2-1-R.  FM acknowledges the support of the French
Agence Nationale de la Recherche (ANR), under the program
ANR-12-BS05-0012 Exo-atmos.

\bibliography{refs}

\begin{deluxetable}{llccc}

\tablewidth{0pt} 
\tabletypesize{\footnotesize} 
\tablecaption{Fitted
  central time and planet-to-star radius ratio of the 67
  transits. \label{tab:results}}

\tablehead{
\colhead{Epoch} & 
\colhead{Filter}   & 
\colhead{$T_C$}  & 
\colhead{$R_p/R_s$} & 
\colhead{Reference}   \\ 
\colhead{}      &
\colhead{}      & 
\colhead{$ (BJD_{TDB} - 2450000)$} &
\colhead{} & 
\colhead{}  }

\startdata

11    & I+z      & $5537.81687       \pm 0.00042$   & $0.1592 \pm 0.0014 $  & (a)       \\ 
22    & I+z      & $5546.76487       \pm 0.00031$   & '' &  ''       \\ 
27    & I+z      & $5550.83227       \pm 0.00030$   &'' &   ''     \\
38    & I+z      & $5559.78035       \pm 0.00020$   &''  & ''      \\ 
43    & I+z      & $5563.84771       \pm 0.00037$   & '' & ''       \\ 
49    & I+z      & $5568.72840        \pm 0.00021$   & '' & ''       \\ 
59    & I+z      & $5576.86382         \pm 0.00027$   &''  & ''       \\ 
65    & I+z      & $5581.74409         \pm 0.00017$   & '' & ''       \\ 
70    & I+z      & $5585.81204         \pm 0.00060$   & '' & ''       \\
97    & i+z      & $5607.77519         \pm 0.00022$   &'' & ''      \\ 
124   & I+z      & $5629.73997         \pm 0.00018$   &'' & ''       \\ 
140   & I+z      & $5642.75453         \pm 0.00020$   & '' & ''       \\ 
141   & I+z      & $5643.56892         \pm 0.00035$   & '' & ''       \\ 
152   & I+z      & $5652.51586         \pm 0.00036$   & '' & ''       \\ 
168   & I+z      & $5665.53204         \pm 0.00033$   & '' & ''       \\ 
173   & I+z      & $5669.59921         \pm 0.00033$   & ''& ''       \\ 
189   & I+z      & $5682.61585         \pm 0.00029$   & '' & ''       \\ 
200   & I+z      & $5691.56382         \pm 0.00025$   & '' & ''       \\ 
211   & I+z      & $5700.51238         \pm 0.00027$   & '' & ''       \\ 
243   & I+z      & $5726.54415         \pm 0.00027$   & '' & ''       \\ 
38    & $r'$        & $5559.78086         \pm 0.00034$   & $0.1619^{+0.0021}_{-0.0023}$ & '' \\ 
76    & $r'$       & $5590.69261         \pm 0.00030$   & '' & '' \\ 
87    & $r'$       & $5599.64053         \pm 0.00033$   & '' & '' \\ 
543   & $R$        & $5970.58521         \pm 0.00039$   & $0.1587^{+0.0049}_{-0.0042}$ &  (b)      \\ 
593   & $R$        & $ 6011.2590               \pm 0.0010 $   & $0.1610^{+0.0030}_{-0.0030}$ &  (f)      \\ 
1032  & clear    & $6368.37461         \pm 0.00087$   & $0.1614 \pm 0.0082 $  & ''  \\ 
499   & $g'$        & $5934.79197         \pm 0.00018$   & $0.1589 \pm 0.0024 $  & (c)       \\ 
499   & $r$        & $5934.79216         \pm 0.00011$   & $0.1578^{+0.0014}_{-0.0016}$ & ''         \\ 
499   & $i'$        & $5934.79224         \pm 0.00012$   & $0.1589^{+0.0015}_{-0.0012}$ & ''         \\ 
499   & $z'$        & $5934.79244         \pm 0.00016$   & $0.1596^{+0.0024}_{-0.0020}$  & ''         \\ 
499   & $J$        & $5934.79270         \pm 0.00048$   & $0.1526\pm0.0048$ & ''         \\ 
499   & $H$        & $5934.79225         \pm 0.00033$   & $0.1524^{+0.0036}_{-0.0034}$ & ''         \\ 
499   & $K$        & $5934.79214         \pm 0.00062$   & $0.1452^{+0.0060}_{-0.0062}$ & ''         \\ 
478   & white    & $5917.70909         \pm 0.00015$   & $0.1591 \pm  0.0018$ & (d)    \\ 
538   & white    & $5966.51756         \pm 0.00045$   & '' & ''   \\ 
554   & white    & $5979.53304         \pm 0.00017$   & '' & ''   \\ 
624   & white    & $6036.47674         \pm 0.00012$   & '' & ''   \\ 
950   & white    & $6301.66875         \pm 0.00008$   & '' & ''   \\ 
1442  & $R$        & $6701.89860         \pm 0.00025$   & $0.1585 \pm  0.0025$ & (e)   \\ 
1469  & $R$        & $6723.86516         \pm 0.00060$   & '' & ''     \\ 
1469  & $i'$        & $6723.86378         \pm 0.00080$   & $0.1724 \pm 0.0030$  & ''    \\ 
1485  & $I$        & $6736.87711         \pm 0.00092$   & $0.1656 \pm 0.0051$ & ''    \\ 
1486  & V        & $6737.69115         \pm 0.00036$   & $0.1607^{+0.0033}_{-0.0037}$ & ''     \\ 
1496  & $i'$        & $6745.82607         \pm 0.00066$   & $0.1724\pm0.003$   & ''    \\ 
1550  & $I$        & $6789.75294         \pm 0.00059$   & $0.1656 \pm0.0051$ & ''   \\ 
1475& $R$        & $ 6728.74251               \pm  0.00049$   & $0.1610^{+0.0030}_{-0.0030}$ &  (f)      \\ 
1480& $R$        & $ 6732.80944               \pm  0.00069$   & '' &  ''      \\ 
1507& $R$        & $ 6754.77381               \pm  0.00025$   & '' & ''       \\ 
1828& $R$        & $ 7015.89832               \pm  0.00042$   & '' & ''       \\ 
1844& $R$        & $ 7028.91467               \pm  0.00037$   & '' &   ''     \\ 
1855& $R$        & $ 7037.86187               \pm  0.00046$   & '' &   ''     \\ 
1860& $R$        & $ 7041.92986               \pm  0.00032$   & '' &   ''     \\ 

1536  & $i'$        & $6778.36469         \pm 0.00035$   & $0.1613^{+0.0021}_{-0.0018}$ & (g)  \\ 
1546  & $i'$        & $6786.49928         \pm 0.00045$   & '' & ''     \\ 
1551  & $i'$        & $6790.56680         \pm 0.00031$   & '' & ''     \\ 
1567  & $i'$        & $6803.58254         \pm 0.00089$   & '' & ''     \\ 
1834  & $i'$        & $7020.78023         \pm 0.00018$   & '' & ''     \\ 
1888  & $i'$        & $7064.70770         \pm 0.00018$   & '' & ''     \\ 
2329  & $i'$        & $7423.45037         \pm 0.00043$   & ''  & ''      \\ 

1536  & $g'$        & $6778.36467         \pm 0.00037$   & $0.1605^{+0.0020}_{-0.0017}$ & (g)  \\ 
1546  & $g'$        & $6786.49990         \pm 0.00025$   & '' & ''     \\ 
1551  & $g'$        & $6790.56728         \pm 0.00051$   & '' & ''     \\ 
1567  & $g'$        & $6803.58268         \pm 0.00105$   & '' & ''     \\ 
1605  & $g'$        & $6834.49439         \pm 0.00049$   & '' & ''     \\ 
1834  & $g'$        & $7020.77958         \pm 0.00040$   & '' & ''     \\ 
1888  & $g'$        & $7064.70746         \pm 0.00031$   & '' & ''     \\ 
1986  & $R$        & $7144.42829         \pm 0.00056$   & $0.1595^{+0.0052}_{-0.0055}$ & ''    \\ 

\enddata \tablecomments{Results of the modelling of 67 transit light
  curves of WASP-43b (see Section \ref{rprs_free}). The reference of
  each transit corresponds to: (a) \cite{Gillon_2012}, (b)
  \cite{Macie_2013}, (c) \cite{Chen_2014}, (d) \cite{Murgas_2014}, (e)
  \cite{Ricci_2015}, (f) \cite{Jiang_2015} and (g) this work. }

\end{deluxetable}

\begin{deluxetable}{lll}
\tabletypesize{\footnotesize}
\tablewidth{0pt} 
\tablecaption{Final results. \label{final_results}}

\tablehead{ \colhead{ Parameter} & \colhead{value} & \colhead{$\pm
    1\sigma$} } \startdata $R_p/R_s$ & 0.15942 & 0.00041 \\ $i$~
(degrees) & 82.11 & 0.10 \\ $a/R_s$ & 4.867 & 0.023 \\ $P$ (d) &
0.813473978 & 0.000000035 \\ $T_0$ ($BJD_{TDB}$) & 2455528.868634 &
0.000046 \enddata \tablecomments{Final results of the joint modelling
  of the light curves (Section \ref{RpRs_simul}) and the timing
  analysis of the transits (Section \ref{timing-a}).}
\end{deluxetable}

\end{document}